\documentstyle[psfig,astrcite,emulateapj]{article}
\def\lsim{\mathrel{\hbox{\rlap{\lower.55ex \hbox {$\sim$}}\kern-.0em
\raise.4ex \hbox{$<$}}}} 
\def\gsim{\mathrel{\hbox{\rlap{\lower.55ex \hbox {$\sim$}}\kern-.0em
\raise.4ex \hbox{$>$}}}} 
\newcommand{\gpm}[3]{$#1^{+#2}_{-#3}$}

\newcommand{\nuc}{\mbox{$\nu_{\rm c}$}}

\newcommand{\grb}{$\gamma$-ray burst}
\newcommand{\grbs}{$\gamma$-ray bursts}

\newcommand{\AVr}{\mbox{$A_{V{\rm r}}$}}
\newcommand{\AV}{\mbox{$A_V$}}
\newcommand{\NH}{\mbox{$N_{\rm H}$}}

\newcommand{\nuV}{\mbox{$\nu_V$}}

\newcommand{\FV}{\mbox{$F_{V0}$}}

\newcommand{\be}[1]{\begin{equation}\label{#1}}
\newcommand{\ee}{\end{equation}}
\newcommand{\msun}{\mbox{M$_\odot$}}

\begin{document}

\title{High column densities and low extinctions of $\gamma$-ray bursts:\\
       Evidence for hypernovae and dust destruction}
\author{Titus J. Galama\altaffilmark{1,2} and Ralph A.M.J. Wijers\altaffilmark{3}}

\altaffiltext{1}{Division of Physics, Mathematics and Astronomy,
 California Institute of Technology, MS~105-24, Pasadena, CA 91125;
 tjg@astro.caltech.edu}
\altaffiltext{2}{Sherman Fairchild Fellow}
\altaffiltext{3}{Department of Physics and Astronomy, State University
                 of New York, Stony Brook, NY 11794-3800, USA;
                 rwijers@astro.sunysb.edu} 

\begin{abstract}
We analyze a complete sample of $\gamma$-ray burst afterglows, and
find X-ray evidence for high column densities of gas around them.
The column densities are in the range $10^{22}-10^{23}$\,cm$^{-2}$, which
is right around the average column density of Galactic giant molecular
clouds.  We also estimate the cloud sizes to be 10--30\,pc, implying
masses $\gsim10^5\,\msun$.  This strongly suggests that $\gamma$-ray
bursts lie within star forming regions, and therefore argues against
neutron star mergers and for collapses of massive stars as their sources.
The optical extinctions, however, are 10--100 times smaller than expected
from the high column densities. This confirms theoretical findings that
the early hard radiation from \grbs\ and their afterglows can destroy the
dust in their environment, thus carving a path for the afterglow light
out of the molecular cloud. Because of the self-created low extinction
and location in star-forming regions, we expect $\gamma$-ray bursts to
provide a relatively unbiased sample of high-redshift star formation. Thus
they may help resolve what is the typical environment of high-redshift
star formation.

\end{abstract} 

          
   \section{Introduction}
       \label{intro}

\grbs\ emit up to 10$^{53}$\,ergs in gamma rays and X rays
in a few seconds, followed by an `afterglow' of X-ray
(\cite{Costa97B}), optical (Van Paradijs et~al.\ \cite*{Paradijs97})
and radio (\cite{Frail97A}) emission that generally lasts days to
months (Van Paradijs, Kouveliotou, \& Wijers \cite*{Paradijs00}).  The
arcsecond localizations of \grbs\ by the detection of these
counterparts have made it possible to study the environments in which
\grbs\ arise. This has provided a number of indications of their
association with massive stars and star formation. First, most \grbs\
lie within the region of UV emission from massive stars
in their host galaxy.
Second, the energies of \grbs\ are comparable to those of
supernovae, further suggesting the deep gravitational collapse of a
few solar masses as the source of energy for \grbs.  Candidates
satisfying these requirements include exploding very massive stars,
also termed hypernovae or collapsars (\cite{Woosley93B,Paczynski98}),
mergers of two neutron stars (\cite{Eichler89}) and the merger of a
neutron star and a black hole (\cite{Mochkovitch93}). Mergers may
be inconsistent with the small offsets between \grbs\ and their hosts
(\cite{Bloom99A,Bulik99}).
Further evidence for the relation between \grbs\ and star formation
has been provided by the fact that the brightness distribution of
\grbs\ agrees well with models in which the \grb\ rate tracks the star
formation rate over the past 15 billion years of cosmic history
(\cite{Totani97,Wijers98,Kommers00A}).  The most direct evidence
relating \grbs\ to a specific type of progenitor has been the
discovery of supernova 1998bw in the error box of GRB\,980425
(\cite{Galama98C}) and the detection of supernova-like light curves
underneath two afterglows, GRB\,980326 (\cite{Bloom99C}) and
GRB\,970228 (\cite{Reichart99A,Galama00A}). Despite
their unusually high luminosity, these supernovae
would often go unnoticed due
to the much brighter \grb\ afterglow; therefore, we do not know whether
most \grbs\ are associated with supernovae or just some of them. Also,
the fact that GRB\,980425 was very nearby and subluminous in gamma
rays by a factor $10^4$ makes it hard to extrapolate its supernova
association to normal \grbs\ without additional considerations.

To elucidate the \grb-supernova association further, we examine
optical and X-ray extinctions of \grb\ afterglows (Sect.~\ref{extin}).
Then we discuss the
evidence these extinctions provide that the majority of long \grbs\
occur in molecular clouds, with dust destruction explaining the unusually
low extinctions (Sect.~\ref{discu}). We thus infer that neutron star
and black-hole mergers are no longer plausible \grb\ progenitors, and 
briefly mention some further implications of our results
(Sect.~\ref{concl}).

%
   \section{Extinction of $\gamma$-ray burst afterglows}
       \label{extin}

      \subsection{Optical extinction}
          \label{extin.optic}

A paradox in associating most \grbs\ with exploding very
massive stars is that one expects the majority of these to lie amidst
highly absorbing molecular clouds. The lack of high restframe visual
extinction has therefore led to considerable skepticism about the
\grb-supernova connection. Here we reinvestigate the extinction for
\grbs\ in a systematic way, by fitting an extinction model to all
afterglows for which the required X-ray and optical data exist. The
model function is 
\be{eq:onepl} F_\nu = \FV [\nu/\nuV]^{-\beta}\exp[-\AVr(1+z)\nu/\nuV], 
\ee 
where $\nuV$ and \FV\ are the observer-frame $V$ band central frequency and
extinction-corrected $V$ band flux, respectively, and \AVr\ is the
rest frame visual extinction; the extinction term is applied
only to optical and infrared data. The results of the fits are given in
table~\ref{ta:onepl}. Note the low optical extinctions. (Some fitted
$\AVr$ are negative, as expected from a fit procedure 
if the values
are less than the fit errors; we did not force $\AV>0$.)  We have used the
simplest possible extinction law here, $A_\nu\propto\nu$.  In metal-rich
environments, the extinction can have features, such as the 2200\AA\
`bump'. We see no significant detection of this bump in any afterglow, 
despite its easy observability at redshifts 1--2, and
therefore neglect it.

The prompt and afterglow X rays are not significantly attenuated,
and there is only one \grb\ in which the search for an optical
afterglow resulted in upper limits well below the expected optical flux
(GRB\,970828; \cite{Groot98A}\footnote{%
   if due to extinction, the factor 300 or more flux deficit requires 
   $\AVr>3$ for this source, which has $z=0.96$ (\cite{Djorgovski00})}.)
Most optical non-detections are adequately explained by adverse observing 
conditions, and consistent with the presence of a normal afterglow. Hence
there is no evidence from non-detections for a significant
population of highly extincted \grbs, or for significant skewing of the
extinction distribution due to the
selection effect that \grbs\ with higher extinction are simply not found.

In the above analysis, we have assumed that the intrinsic spectrum of
the afterglow is a pure power law. The theory of afterglows allows
a so-called `cooling break' in the spectrum between 
X rays and optical, where the power-law slope steepens by 1/2.
A comparison of the mean optical-to-X-ray slope, $\beta_{OX}$, and the X-ray
spectral slope, $\beta_X$, shows that they are indeed different, indicating
that such a break may be present: 
$<\!\beta_X\!-\!\beta_{OX}\!>=\!0.31 \pm 0.05$. 
Therefore we redo our fits with an extra break. The fit function is 
\begin{eqnarray}
   \label{eq:twopl}
   F_\nu & = & \FV [\nu/\nuV]^{-(\beta +
1/2)}\exp[-\AVr(1+z)\nu/\nuV]\,\nonumber\\
& \times &  \left(\frac{[\nu(1+z)/\nuc]^{-n/2} + 1}{ [\nuV(1+z)/\nuc]^{-n/2} +
1}\right)^{-1/n}. 
\end{eqnarray}
The index $n$ controls the sharpness of the break. We use $n=2$, which
provides a smooth transition over 2 dex in frequency, and find little
change in the fits if we vary the width in the range 1--3 dex.
The extra datum needed to
constrain this fit is $\beta_X$.  To treat this uniformly with the
other data, we add an artificial data point with ten times greater
frequency than the first X-ray point, and give it a value and error so
that a fit to the two X-ray points gives the correct values and errors
of the X-ray flux and spectral slope.  The results of
the new fits are also given in table~\ref{ta:onepl}, showing hardly any
change in the extinctions.
      
\subsection{X-ray absorption}
\label{extin.xray}

\begin{figure}[h!]
  \psfig{figure=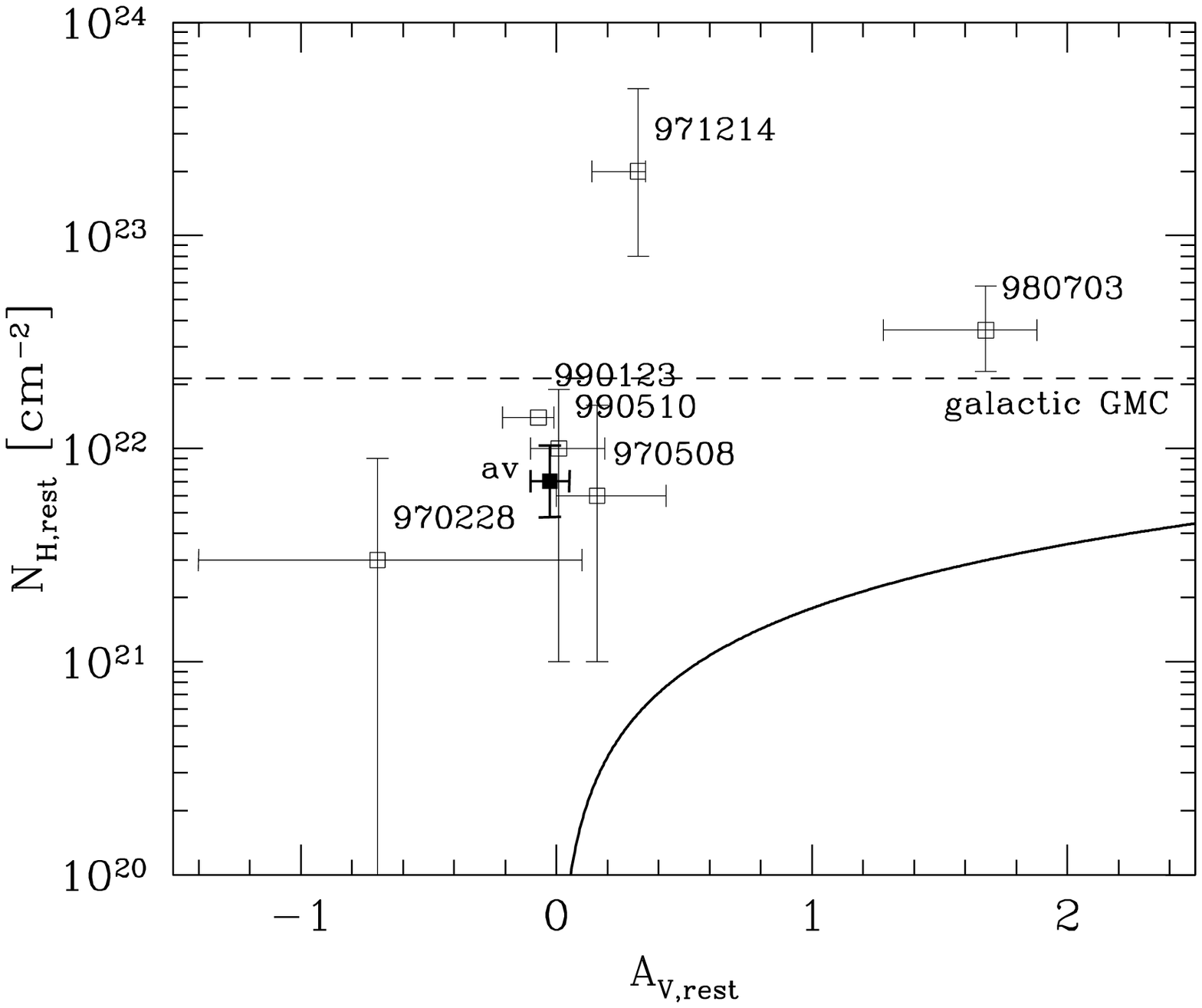,width=\columnwidth}
  \caption[]{\footnotesize The hydrogen column density \NH\ versus the optical
      extinction at V band \AVr\ for a number of \grb\ afterglows; data
      are from table~\ref{ta:onepl}.  The solid curve is the Galactic
      \AV-\NH\ relation  (\cite{Predehl95}). The solid symbol is the weighted 
      mean of the four lowest points, assuming 50\% for the unknown error of
      \NH\ for GRB\,990123. The dashed line shows the average column density
      of a giant molecular cloud of 170\msun/pc$^2$ (\cite{Solomon87}).
     \label{fi:avnh}
            }
\end{figure}

Our results confirm in a more rigorous way the casual impression of
low optical extinction in \grbs\ and therefore seem to contradict the
notion of exploding massive stars being the progenitors of
\grbs. However, the hydrogen column densities towards the \grbs\
reveal startlingly contrasting evidence. In table~\ref{ta:onepl} we
collect from the literature values of the hydrogen column density
towards \grbs\ as derived from soft X-ray absorption. To obtain these
values, we subtracted the Galactic foreground column (\cite{Dickey90})
from the total
measured value. Then we accounted for the fact that the soft X-ray
optical depth is a strong function of energy, $\tau_X\propto E^{-2.6}$
(\cite{Morrison83}), which implies that we have to multiply the
foreground-subtracted column by $(1+z)^{2.6}$ for a source at redshift
$z$, to get the true column density in the rest frame of the
$\gamma$-ray burst. (Note that while in principle the photoelectric
absorption has richer structure due to absorption edges of individual
atomic species, the presently available spectral resolution and S/N
does not allow this to be discerned.)  This is a
large correction, ranging from 4 to 50 among the sample, and the
resulting restframe hydrogen column densities are in the range
$10^{21.5}-10^{23.3}$\,cm$^{-2}$. To emphasize the contrast between the
optical and X-ray results, we show in Figure~\ref{fi:avnh} the
correlation between optical and X-ray extinction. The solid curve
indicates the relation between $A_V$ and $N_H$ for the Milky Way
(\cite{Predehl95}), so we see that the observed visual extinctions are
10--100 times smaller than expected for the observed X-ray
absorption. 

Both types of extinction are due to heavy elements, so
metallicity differences cannot change the ratio. However, the
optical/UV extinction is due to dust grains, whereas the X-ray
extinction is due to K- and L-shell electrons of intermediate-mass
elements (mostly C and O) and therefore does not
depend on whether the atoms are in a gas or a solid.  The X-ray
extinction is therefore a better measure of the total column
density. However, converting the X-ray absorption to a hydrogen column
density, as is customary, {\it does\/} depend on metallicity.  Since
the high-redshift regions we are probing may have lower metallicities,
the true column densities can only be larger than those we have
derived.  (Since the regions are very actively star forming, their
metallicity may not be much less than in the Milky Way, though.)

   \section{Implications for $\gamma$-ray burst environments}
       \label{discu}

The values of the X-ray column densities are very high, and typical
of the column densities through giant molecular clouds (Fig.~\ref{fi:avnh};
\cite{Solomon87}). This strongly suggests that most \grbs\ are located
in molecular clouds. We now attempt to constrain the size of the clouds, and
thereby their mass.

First, the surprisingly low extinctions may be explained by recently 
proposed dust destruction: dust is sensitive to the UV
and X-ray radiation from the \grb\ and its afterglow. UV and
X-ray light heats the grains, and out to about 20\,pc can evaporate them
(\cite{Waxman00,Fruchter00B}).  Waxman and Draine find that a prompt
flash like that seen only in GRB\,990123 is needed to muster enough UV
light; however, Fruchter et~al.\ show that the X-ray flux is at least
as efficient in heating the grains. Since these come mostly from the 
prompt burst emission,  the uncertain UV flashes are not needed 
in their model. They find that
average \grbs\ can evaporate all dust out to 20\,pc. Beyond this, dust may
be shattered by strong grain charging. Since the effect of this on dust
extinction properties is unclear, we shall not consider it here.
From our low optical extinctions, we conclude that dust destruction must
be taking place. This limits the bulk of the cloud to lie within about
20\,pc of the \grb.

A definite lower limit to the size of the absorbing region follows from the
fact that the afterglow radius after a day is about 0.1\,pc (e.g.,
\cite{Wijers99B}). This firmly excludes any remains of the exploded star
as the source of X-ray or optical absorption.
The size of the absorbing cloud can also be bounded from below using absorption
lines of MgI in the spectra of many optical transients
(e.g., \cite{Vreeswijk00,Metzger97B}). Afterglows tend to have
strong MgI lines, especially relative to MgII, indicating they originate
in denser regions than the normal diffuse ISM. We therefore suppose they
originate in the same region that causes the large X-ray columns. The fact
that this MgI is still visible after a day means it has not all been ionized
away. MgI has an
ionization energy of 7.84\,eV, and photons above 13.6\,eV are
stopped by H very near the \grb.  Averaging the ionization
cross section over this energy range, weighted by the afterglow fluence
spectrum, we find an optical depth to MgI ionizing photons of 0.7$N_{\rm
H22.5}$ (assuming the solar value of [Mg/H]). This means that for most
of the column density range, the optically thin limit is adequate for
judging the survival of MgI at the edge of the cloud. Recombination times
are much too long to play a role.

Integrating the \grb\ flux over time and over the same energy range, weighted
by the energy-dependent cross section, we find that an average Mg atom
would intercept 
$2.1\times10^4E_{52}^{4/3}n^{1/2}\epsilon_{B,-2}^{5/6}\epsilon_{\rm e,-1}^{4/3}(d/1\,{\rm pc})^{-2}$ 
ionizing photons
(for a typical afterglow with $\beta=0.75$, where 
$\epsilon_{B}=0.01\epsilon_{B,-2}$ and  
$\epsilon_{\rm e}=0.1\epsilon_{\rm e,-1}$ are the equipartition fractions in
the notation of Wijers \& Galama 1999\nocite{Wijers99B}).
Therefore, some of the Mg must be many parsecs from the \grb\ in order
to survive. The Mg lines are usually rather saturated, so we can only
get a lower limit to the MgI column density.
For the case of GRB\,990510 (\cite{Vreeswijk00}), this limit is
$10^{13.3}$\,cm$^{-2}$, which for normal Mg abundance is contained in less
than
$10^{21}$\,cm$^{-2}$ of total \NH, accounting for depletion.
This is less than the total observed X-ray column, so
only a fraction of Mg need remain neutral.
To set a conservative lower limit, we shall tolerate 10 ionizations for the
average Mg atom at the edge, implying survival of less than
${\rm e}^{-10}=5\times10^{-5}$ of Mg anywhere in the cloud.
Then the lower limit to the cloud size becomes
45$\,E_{52}^{2/3}n^{1/4}\epsilon_{B,-2}^{5/12}\epsilon_{\rm e,-1}^{2/3}$\,pc.
This is quite sensitive to burst parameters, and often above the upper limit
from dust destruction. Therefore, Mg lines need not occur in every burst, and
their presence should be correlated with burst strength.

Together,
dust destruction and MgI survival constrain the size of the
cloud to be tens of parsecs.  The cloud therefore has a density of
$500\,N_{\rm H22.5}/R_{20}$\,cm$^{-3}$ and a mass of $4\times10^5\,N_{\rm
H22.5}R_{20}^2\,\msun$ (where $N_{\rm H22.5}= \NH/10^{22.5}$\,cm$^{-2}$ and
$R_{20}=R/20$\,pc).   These parameters are very much like those of
giant molecular clouds (\cite{Solomon80,Solomon87}).  We therefore
consider our findings strong evidence that almost all (long) \grbs\
are associated with giant molecular clouds, and therefore with
star-forming regions. This, in turn, speaks in favor of massive stars
rather than compact-object mergers as the progenitors of \grbs.

Further predictions of the location of \grbs\ in molecular clouds
are associated with absorption/scattering processes
of the \grb\ emission in the cloud: (i) Absorption and reradiation
by sublimating dust in the infrared may produce a reradiation echo
with a thermal spectrum that peaks in restframe infrared on a time
scale of several tens of days (\cite{Waxman00}). (ii) Scattering of
the afterglow's light by dust outside the dust-vacated region may
produce a scattering echo on time scales of tens to hundreds of days
(\cite{Esin00}). This echo has a spectrum similar to that of the afterglow.
Each echo can emit 10$^{41-42}$
erg/s. (iii) The far UV radiation will be absorbed by H$_2$, causing
a strong drop in the UV at 1650\AA\, and 1300 \AA, and fluorescence
will result in restframe UV emission on time scales of days to months
(\cite{Draine00}). (iv) If burst radiation is collimated, it is likely that
the later, softer emission is less collimated, enabling us to see more
afterglows than \grbs\ (e.g., \cite{Rhoads97B}).  However, because dust is
destroyed only along the collimated path of the initial hard radiation,
such `\grb-less afterglows' would not be visible in optical and near-IR
from embedded sources. Only in far-IR, mm, and radio could the frequency
of afterglows be significantly greater than that of \grbs.

Our findings may also have some indirect bearing on the issue of the
cosmic star formation history, in the following sense: there has been much
recent debate on the relative importance of UV and far-IR radiation in
counting the star formation rate at high redshift 
(e.g., \cite{Madau98,Barger99}). In both cases, one 
counts the location of massive, UV-producing stars, but in the far-IR case
it is found/assumed that the majority of these are deeply shrouded in dust,
concentrated in ultraluminous IR galaxies (ULIGs, e.g.\ \cite{Sanders96}).
Since \grb\ radiation escapes fairly well even from ULIGs, \grb\ locations
might provide an unbiased sample of massive-star
locations at redshifts 1--4. This means that a far-IR study of \grb\ host
galaxies should help resolve the issue of what type of host, ULIGs or
UV-emitting smaller galaxies, are the dominant source of massive-star
production at these redshifts.

%
    \section{Conclusions}
        \label{concl}

We have examined a complete sample of \grb\ afterglows, namely those
with known redshift and X-ray column density, for which
optical to X-ray data allow a determination of reddening. As a sample, they
provide strong evidence for high X-ray column densities, without a single
good exception.  However, the individual measurements in a given source
are very significant only for GRB\,980703 and GRB\,980329. Therefore,
good X-ray spectra are required for more sources in order to confirm
our findings and pin down the parameters of the clouds better. The size
and origin of the absorbing matter are constrained by the low extinction,
the blast wave size, and the survival of MgI. High-resolution measurements
of the Mg absorption lines to better determine
the location and column density of the Mg absorber
are needed to strengthen the lower limits on cloud size and mass.

In short, we find high X-ray column densities and low optical extinctions
for \grb\ afterglows, from which we infer that
(i) Most \grbs\ are embedded in large molecular clouds.
(ii) \grbs\ are therefore likely produced by dying massive stars, and not by
      mergers of neutron stars and/or black holes.
(iii) The low optical extinctions of \grbs\ confirm theoretically
      predicted dust destruction by their hard radiation, which `paves the
      way' for the optical afterglow to escape even large clouds.
(iv) \grb\ host studies may help identify the dominant sources of high-redshift
star formation.

\medskip
We are grateful to Shri Kulkarni,  Daniel Reichart, James Rhoads, and 
    Phil Solomon for
   useful discussions. TJG is supported by the Sherman Fairchild
   Foundation.



\twocolumn

\begin{table*}[h!]
{\footnotesize
\caption[]{\footnotesize The results of fitting a single power law
plus dust absorption (eq.~\ref{eq:onepl})
and the results of fitting a broken power law plus dust
absorption (eq.~\ref{eq:twopl}) to
the X-ray to optical/infrared spectral flux distributions at the given
epoch of \grb\ afterglows (see text for fits and parameter and sample
definitions). \nuc\ and \AV\ are in the rest frame if $z$ is known.
References and notes: $\beta_{\rm X}$ and N$_{\rm H}$ are from Owens et al.\
(1998)\nocite{Owens98A}, unless stated otherwise; 970228
(\cite{Galama00A,Djorgovski99A}); 970508
(\cite{Galama98D,Galama98B,Bloom98C}); 971214
(\cite{Ramaprakash98A,Halpern98A,Wijers99B,Kulkarni98B}; in the fit we
excluded the infrared data because of the observed spectral bump
\cite{Ramaprakash98A}); 980329 (\cite{Reichart99B,Zand98B}); 980519
(\cite{Halpern99A}); 980703 (\cite{Vreeswijk99A} [$\beta_{\rm X}$,
N$_{\rm H}$], \cite{Djorgovski98F}); 990123 (\cite{Piro00} [Fig. 6,
$\beta_{\rm X}$], \cite{Galama99A,Heise99A} [N$_{\rm H}$]); 990510
(\cite{Harrison99,Stanek99,Kuulkers00,Vreeswijk99C}).
\label{ta:onepl} }
\newcommand{\zmin}{\makebox[0pt][r]{$-$}}
\renewcommand{\arraystretch}{1.3}
\begin{tabular*}{\textwidth}{@{}l@{\extracolsep{\fill}}l@{\hspace{2pt}}|@{}l@{}l@{\hspace{2pt}}l@{}l@{\hspace{2pt}}|l@{}l@{\hspace{2pt}}l@{}l@{}l@{\hspace{2pt}}|l@{}l@{}l@{}}
\hline
~ & ~ & \multicolumn{4}{c|}{single power law plus dust} 
      & \multicolumn{5}{c|}{broken power law plus dust} & ~ & ~ & ~ \\
\hline
GRB &  Epoch & log(\FV) & $\beta$ & \AVr & $\chi^{2}/$ & log(\FV) 
& $\beta$ & \AVr & log\,\nuc & $\chi^{2}/$ & $\beta_{\rm
X}$ & N$_{\rm H}$ & $z$ \\ 
    & (UT) & (Jansky) & & & dof & (Jansky) & & & & dof & & ($\times10^{21}$) & \\ \hline \hline

970228 & 0228.99 & \gpm{-5.2}{0.4}{0.5} & \gpm{0.63}{0.13}{0.13} &
\gpm{\zmin 0.8}{0.7}{0.7} & 0.3/1 & \gpm{-5.1}{0.5}{0.6} &
\gpm{0.46}{0.29}{0.14} & \gpm{\zmin 0.7}{0.7}{0.8} & & 0.3/1 &
\gpm{0.96}{0.19}{0.19} & ~~\gpm{3}{6}{4} & 0.695\\

970508 & 0510.98 & \gpm{-4.50}{0.12}{0.11} & \gpm{0.84}{0.06}{0.05} &
\gpm{0.02}{0.15}{0.14} & 2.2/3 & \gpm{-4.38}{0.14}{0.10} &
\gpm{0.50}{0.35}{0.20} & \gpm{0.16}{0.16}{0.27} & & 2.1/3 &
\gpm{0.99}{0.29}{0.07} & ~~\gpm{6}{10}{5} & 0.835\\


971214 & 1215.50 & \gpm{-4.72}{0.14}{0.13} & \gpm{0.90}{0.05}{0.05} &
\gpm{0.38}{0.11}{0.09} & 0.2/1 & \gpm{-4.81}{0.40}{0.08} &
\gpm{0.38}{0.60}{0.03} & \gpm{0.32}{0.18}{0.03} & & 1.8/1 &
\gpm{1.03}{0.51}{0.22} & \gpm{200}{290}{120} & 3.42\\

980703 & 0704.40 & \gpm{-4.09}{0.07}{0.06} & \gpm{1.05}{0.03}{0.03} &
\gpm{1.63}{0.20}{0.20} & 2.0/1 & \gpm{-4.06}{0.23}{0.07} &
\gpm{1.01}{0.06}{0.25} & \gpm{1.68}{0.40}{0.20} & & 2.1/1 &
\gpm{18}{2}{3} & ~\gpm{36}{22}{13} & 0.966\\

990123 & 0124.65 & \gpm{-4.94}{0.06}{0.06} & \gpm{0.66}{0.03}{0.03} &
\gpm{\zmin 0.20}{0.17}{0.16} & 2.0/5 & \gpm{-4.93}{0.22}{0.06} &
\gpm{0.65}{0.04}{0.15} & \gpm{\zmin 0.07}{0.14}{0.06} &
\gpm{18.8}{0.6}{3.4} & 1.9/5 & \gpm{0.97}{0.04}{0.04} & ~14 & 1.60 \\

990510 & 0511.26 & \gpm{-4.20}{0.08}{0.09} & \gpm{0.86}{0.04}{0.04} &
\gpm{\zmin 0.22}{0.23}{0.23} & 0.15/2 & \gpm{-4.09}{0.11}{0.18} &
\gpm{0.60}{0.25}{0.20} & \gpm{0.01}{0.11}{0.18} & \gpm{16.7}{3.2}{1.7} & 0.0/3 &
\gpm{1.12}{0.12}{0.12} & ~\gpm{10}{9}{9} & 1.62 \\

\hline 980329 & 0402.99 & \gpm{-5.70}{0.17}{0.20} &
\gpm{0.91}{0.10}{0.07} & \gpm{2.9}{1.1}{0.8} & 0.0/0 &
\gpm{-5.65}{0.30}{0.22} & \gpm{0.85}{0.15}{0.20} & \gpm{3.0}{1.3}{0.9} &
& 0.4/0 & \gpm{1.63}{0.41}{0.36} & ~\gpm{10}{8}{4} & \\

980519 & 0520.34 & \gpm{-4.34}{0.16}{0.15} & \gpm{0.98}{0.06}{0.06} &
\gpm{0.8}{0.4}{0.4} & 0.2/2 & \gpm{-4.31}{0.26}{0.17} &
\gpm{0.92}{0.11}{0.27} & \gpm{0.9}{0.6}{0.4} & & 0.4/2 &
\gpm{1.52}{0.70}{0.57} & ~~\gpm{4}{10}{5} & \\ \hline

\end{tabular*}
}
\end{table*}

\end{document}